# Polariton microfluidics for nonreciprocal dragging and reconfigurable shaping of polaritons

Zhenyang Cui[1,2,3], Sihao Xia[1,2,3], Lian Shen[1,2,3], Bin Zheng[1,2,3], Hongsheng Chen[1,2,3]\*, Yingjie Wu[1,2,3]\*

1 ZJU-Hangzhou Global Scientific and Technological Innovation Center, Zhejiang University, Hangzhou 310027, China.

2 International Joint Innovation Centre, Key Lab. of Advanced Micro/Nano Electronic Devices & Smart Systems of Zhejiang, The Electromagnetics Academy at Zhejiang University, Zhejiang University, Haining 314400, China.

3 Jinhua Institute of Zhejiang University, Zhejiang University, Jinhua 321099, China.

\*Correspondence to: hansomchen@zju.edu.cn (H.C.); yingjie.wu@zju.edu.cn (Y.W.).

**Abstract**

Dielectric environment engineering is an efficient and general approach to manipulating polaritons. Liquids serving as surrounding media of polaritons have been used to shift polariton dispersions and tailor polariton wavefronts. However, those liquid-based methods have so far been limited to their static states, not fully unleashing the promises offered by the mobility of liquids. Here, we propose a microfluidic strategy for polariton manipulation by merging polaritonics with microfluidics. The diffusion of fluids causes gradient refractive indices over microchannels, which breaks the symmetry of polariton dispersions and realizes the non-reciprocal dragging of polaritons. Based on polariton microfluidics, we also design a set of on-chip polaritonic elements to actively shape polaritons, including planar lenses, off-axis lenses, Janus lenses, bends, and splitters. Our strategy expands the toolkit for the manipulation of polaritons at the subwavelength scale and possesses potential in the fields of polariton biochemistry and molecular sensing.

**Introduction**

Polaritons are coupled photon and material excitation waves[1, 2]. The strongly confined light field and highly enhanced field intensity render polaritons a powerful tool for light control at the subwavelength scale[3]. Versatile strategies have been developed to manipulate polaritons[4], among which dielectric environment engineering attracts attention due to its easy experimental implementation, flexible choices of components, and broad applicability to polariton modes. The dielectric environment control of polaritons could be used to modify dispersions[5, 6], reduce damping[7, 8], shape wavefronts[9-11], or offer polaritons additional active tuning knobs, such as temperature[12], laser pulse[13], and bias voltage[14].

Liquids with various dielectric permittivities are promising candidates for the dielectric environment engineering of polaritons[15]. It has been reported that replacing air with liquids can shift polariton dispersions to higher momenta[16, 17], which could be used for liquid-phase in situ molecular sensing[18]. Based on such dispersion mismatch, one could build air/liquids interfaces with proper geometries to bend polariton wavefronts[19]. However, those studies just used a single liquid in its static state and did not fully unleash the potential of liquids in the field of polaritonics. Liquids have been extensively employed in optofluidics, the convergence of optics and microfluidics[20-22], to create gradient refractive indices by diffusion[23-26]. Such gradients can be dynamically controlled by optimizing microchannel geometries, liquid components, or flow rates. This methodology is expected to be translated into the polaritonic regime, providing novel opportunities for the control of polaritons which is hardly attainable in solid-state and static-liquid-state systems.

Here, we propose a microfluidic strategy for the manipulation of polaritons by fusing polaritonics with microfluidics. Rooted in the precise control of dielectric permittivity gradients and distributions, our method can break the symmetry of in-plane wavevectors and serve as a more universal platform to support the nonreciprocal dragging of polaritons beyond electron drifting in graphene. Besides, through optimizing microchannels, we also realize the reconfigurable shaping of polaritons and design polaritonic lenses, bends, and splitters as essential elements of polariton circuits. Our results not only pave a microfluidic way to on-chip light flow control at the subwavelength scale but also find potential applications in molecular sensing, particle trapping, and biological nanoimaging.

**Results**

*Nonreciprocity enabled by dielectric environment gradients*

The in-plane momentum ($q$) of polaritons is highly sensitive to the dielectric environment, namely, the dielectric permittivities of the superstrate ($\varepsilon_2$) and substrate ($\varepsilon_0$). For convenience, we fix the superstrate as air ($\varepsilon_2 = 1$) and just consider the impact of substrates. Taking phonon polaritons (PhPs) in hexagonal boron nitride (hBN) as an example (inset of Fig. 1a), their in-plane momentum as a function of $\varepsilon_0$ is displayed in Fig. 1a (see Section S1 for calculations), where the refractive index $n_0 = \sqrt{\varepsilon_0}$ is also provided. Figure 1b shows the corresponding isofrequency surface (red surface), revealing the $\varepsilon_0$-dependent wave vectors ($k_j$, $j = x, y$). Due to the in-plane isotropic dispersion of PhPs in hBN, the contour line at a given $\varepsilon_0$ is a circle, as confirmed by the gray circle in Fig. 1c which represents the intersections between the red surface and gray slice at $\varepsilon_0 = 2$ in Fig. 1b.

If we consider a slice in the *k*-space with gradually changed $\varepsilon_0$ ($\partial\varepsilon_0/\partial k_{x,y} \neq 0$), for example, $\partial\varepsilon_0/\partial k_x = -1.86$ (blue surface in Fig. 1b), the obtained isofrequency contour transitions into anisotropic, as shown in the blue ellipse in Fig. 1c. Such asymmetric isofrequency contour is similar to that of plasmon polaritons (PPs) on drift-biased graphene (known as Fizeau drag)[27-31], implying that the dielectric environment gradients could break the symmetry of polariton propagation and give rise to the nonreciprocal dragging of polaritons. As far as we know, bias-induced Fizeau drag is currently limited to graphene PPs due to the comparable electron drift velocity ($v_d \approx 3\times10^5$ m s$^{-1}$) and polariton group velocity ($v_p \approx 2\times10^6$ m s$^{-1}$) in graphene[31]. By contrast, the microfluidic method could overcome this limitation because the dragging effect is driven by refractive index gradients which are also applicable to other polariton modes.

*Dragging polaritons by flowing liquids*

In order to create gradually changed $\varepsilon_0$ for polariton dragging, we propose a microfluidic strategy by merging polaritonics with microfluidics. We first design a microfluidic structure composed of a circular chamber, two inlets (Inlet 1, Inlet 2), and two outlets (Outlet 1, Outlet 2), as sketched in Fig. 2a (see Section S2 for the details of models). Two miscible liquids with different refractive indices are injected into the chamber made of polydimethylsiloxane (PDMS) at a certain flow rate ($Q$). Here we choose aniline ($n_{\text{aniline}}$ = 1.586) and water ($n_{\text{H}_2\text{O}}$ = 1.332)[15]. Their diffusion leads to concentration gradients over the cross-section of the chamber, $0 \leq C(x, y) \leq 1$, normalized to the concentration of aniline, as shown in Fig. 2b at $Q = 5\times10^{-4}$ m s$^{-1}$. The refractive index distribution can be calculated by $n(x, y) = \sum C_m(x, y) n_m$, $m = 1, 2$ (ref [23]). Figure 2c shows the obtained refractive index. The line scan profiles across the chamber center are displayed in Fig. 2d, indicating a gradually increased refractive index towards Inlet 1 (the $-x$ half space) and a gradually decreased refractive index towards Inlet 2 (the $+x$ half space). A larger flow rate leads to a steeper gradient, whereas a smaller flow rate forms a gentler gradient in the $x$–$y$ plane. The concentration and refractive index distribution are uniform along the $z$ direction, as seen in Fig. S2 in Supporting Information.

The hBN disc with a thickness of 100 nm is then put over the chamber. A vertically polarized dipole above the center of the hBN disc is used to excite PhPs. The in-plane momentum of PhPs differs when using aniline and water as substrates (Fig. 1a). The dielectric permittivity distinction between aniline and water (blue surface in Fig. 1b) is theoretically sufficient for polariton dragging assuming a linear gradient in the $k$-space, as indicated by the green ellipse in Fig. 1c. Figure 2d indicates that the refractive index gradients introduced by the diffusion between aniline and water could create asymmetric isofrequency contours, although thses gradienst are not strictly linear. Indeed, the simulated field distribution in Fig. 2e denoted by the real part of the electric field in the $z$-direction, Re($E_z$), shows the asymmetric and nonreciprocal propagation of PhPs over microfluids. An analogue polaritonic Doppler effect can be found from the extracted line scan profiles in Fig. 2f: PhPs have a shorter wavelength ($\lambda_p$) when propagating along the positive refractive index gradients, whereas PhPs that propagate along the negative refractive index gradients have a longer wavelength. Figure 2f also indicates that such nonreciprocity can be actively controlled by the flow rate: A larger $Q$ leads to quickly saturated refractive indices and a higher dragging efficiency.

Due to the nonlinear gradients, the polariton wavelength shift ($\Delta\lambda_p$) in our microfluidic system is nonuniform, which is different from the drift-induced nonreciprocal graphene PPs. To quantitatively evaluate the dragging efficiency of our strategy, the averaged polariton wavelength shifts along the $+x$ ($\overline{\Delta\lambda_p^+}$) and $-x$ directions ($\overline{\Delta\lambda_p^-}$) at $Q = 5 \times 10^{-3}$ m s$^{-1}$ are calculated and normalized to the polariton wavelength ($\lambda_{p0}$ = 2.06 μm) in the absence of gradients, $n(x, y) = (n_{H_2O} + n_{\text{aniline}})/2$. The obtained dragging efficiency along the $+x$ direction $\overline{\Delta\lambda_p^+}/\lambda_{p0}$ is 10%, whereas $\overline{\Delta\lambda_p^-}/\lambda_{p0} = -9\%$ along the $-x$ directions, comparable to the dragging efficiency of PPs in graphene achieved by current bias (2%~8%)[30, 31].

To analyze the nonreciprocal dragging in the frequency domain, we prerformed the Fourier transform (FT) of the electric field in Fig. 2e. The absolute value of FT is shown in Fig. 2g, indicating asymmetric isofrequency contours between the inner (at $n_0 = 1.332$) and the outer (at $n_0 = 1.586$) circular IFCs. The dispersion diagrams with and without refractive index gradients are compared in Fig. 2h, where the in-plane momenta of the nonreciprocal polaritons are calculated using the averaged polariton wavelengths, $q = 2\pi/\bar{\lambda}_p$. Different from the symmetric branches for the uniform mixture, asymmetry can be observed with refractive index gradients: a positive gradient can depress the dispersion and a negative one can lift the dispersion.

*Polaritonic lenses enabled by microfluidics*

In the low Reynolds number regime, the diffusion between different liquids creates smooth interfaces with gradually changed refractive indices[20]. This phenomenon could be used to shape the wavefront of polaritons and build polaritonic lenses to focus light beyond the diffraction limit. As shown in Fig. 3a, a microfluidic system with a core flow and two cladding flows is designed for polariton shaping and focusing. The concentration distribution and refractive index profile over the rectangular chamber are displayed respectively in Fig. 3b and Fig. 3c, where the flow rates of the core ($Q_{co}$) and cladding flows ($Q_{cl1}$, $Q_{cl2}$) satisfy $Q_{co} = Q_{cl1} = Q_{cl2} = 5\times10^{-4}$ m s$^{-1}$.

When PhPs launched by the left edge of the hBN flake travel over the rectangular chamber, the wave vectors and Poynting vectors bend towards the center of the streamline, as seen from the simulated field distributions in Figs. 3d,e. Such an effect gives rise to a focal spot at a certain distance away from the chamber edge, which is defined as the focal length of the polaritonic lens (*f*). The extracted field intensity at *f* = 19.5 μm is displayed in Fig. 3j (red curve), indicating an enhanced field intensity. The focusing resolution denoted by the full width at half maximum (FWHM) of the Lorentzian fitting is 1.58 μm, which is approximately one fifth of the free-space wavelength. Besides, the focusing effect can be adjusted by the flow rate of microfluids: increasing the flow rate leads to, in principle, a shorter focal length and an enhanced resolution (see details in Fig. S3).

In the above case, microflows are symmetric to the *x* direction because $Q_{cl1} = Q_{cl2}$, making the focal line always along the *x* axis. When $Q_{cl1} \neq Q_{cl2}$, microflows would be asymmetric, so would the associated refractive index gradients, which could result in an additional lateral shift of the focal line, known as off-axis focusing[32, 33]. To explore this effect, we set $Q_{co} = 10Q_{cl1} = Q_{cl2} = 5\times10^{-4}$ m s$^{-1}$ as an example. In Figs. 3f,g one can find that the concentration distribution and refractive index profile bend towards the side with the lower flow rate. Accordingly, asymmetric field distributions are observed in Figs. 3h,i. The obtained focal spot is located at the −*y* axis with an off-axis angle (*θ*) of 10.3° to the *x* direction. The obliquity of the focal line can be tuned by controlling the flow rate ratio, $Q_{cl2}/Q_{cl1}$: a larger flow rate ratio generally leads to an enhanced obliquity, as shown in Fig. 3k (more results can be found in Fig. S4).

As the refractive index distribution along the direction normal to polariton propagation dominates the bending direction of polaritons, an opposite effect could be achieved by exchanging the core and cladding flows. A polaritonic Janus lens that exhibits distinct polaritonic responses along opposite directions could be realized at the interface where the refractive index distribution flips[34]. To do this, in Fig. 4a, we design a microfluidic system with symmetric structures but exchanged core and cladding flows at the two sides. The anti-symmetric concentration and refractive index distributions with respect to the chamber center can be seen in Figs. S5a,b in Supporting Information. A gold strip antenna illuminated by an *x*-polarized plane wave is put upon the center of the hBN flake to launch polaritons propagating along opposite directions. The obtained field distributions are displayed in Figs. 4b,c. In the −*x* direction, a focal spot at the position of *f* is observed, akin to the focusing effect in Fig. 3. By contrast, in the +*x* direction, the wave vectors and Poynting vectors bend away from the *x* axis, leading to a diverging effect. Figure 4c shows the extracted field distribution along the blue and green dashed lines at *f* = *f'* = 30.5 μm, from which the enhanced and weakened intensities can be observed respectively at the −*x* and +*x* directions, corroborating the realization of the focusing and diverging effects at the same time in our polaritonic Janus lens.

*Transformation polaritonics with microfluidics*

The refractive index gradients introduced and tailored by microfluidics further allows us to shape polaritons by means of transformation polaritonics[35-38], the fusion of polaritonics and transformation optics. We first propose a polaritonic bend to rotate the direction of polariton propagation by 180° using the microfluidic system in Fig. 4d with a semicircular microchannel. The refractive index of the bend should satisfy $n_0 = A/r$, where *A* is an arbitrary constant and *r* is the distance to the center of the semicircular channel[39]. Such refractive indices could be achieved by controlling the flow rates of microflows, as shown in Fig. S6e in Supporting Information. The electric field overlaid on the schematic in Fig. 4d confirms that PhPs smoothly travel through the bend with nearly no reflection, despite the intrinsic loss of polaritons (see details in Fig. S6). Polaritonic bends with arbitrary bend angles could be straightforwardly achieved by adjusting the microchannel geometries.

Following a similar transformation polaritonics concept, a polaritonic splitter could be achieved by merging two bends together[40]. As an example, we designed a Y-shaped microchannel with a split angle ($\varphi$) of 37.5° in Fig. 4e. The simulation result (overlaid on the schematic) shows that the incoming polariton beam splits into two branches because of the gradient refractive index distribution. The field intensities extracted along the directions normal to polariton propagation are comparable in the two branches (see details in Fig. S7 in Supporting Information).

**Discussion**

For the sake of convenience, we just consider the ideal configuration in our simulations. For example, hBN flakes are suspended over the microfluidic channels. In practical situations, hBN flakes might be supported by a PDMS layer above the fluids used to seal the channels. We note that, the existence of a thin PDMS layer will not weaken our main conclusions because the refractive index gradients still exist (see details in Fig. S8 in Supporting Information). The microfluidic strategy could also be used to shape PPs on graphene, as seen in Fig. S9 in Supporting Information. More fluids with larger refractive indices can also be used to create a sharper refractive index gradient, for example, high-index oil ($n$ = 1.7)[26]. Besides, more delicate channels with varied widths and depths could add more degrees of freedom to polariton manipulation with microfluidics.

**Conclusion**

In summary, we propose a microfluidic strategy to manipulate polaritons rooted in the dielectric environment engineering of polaritons. The diffusion of liquids creates gradient refractive index distribution over microchannels which breaks the symmetry of polariton dispersion and gives rise to non-reciprocal dragging of polaritons. We further show the ability of microfluidics to shape the wavefronts of polaritons, yielding multifunctional polaritonic lenses with subdiffractional resolution. Polaritonic bends and splitters are also conceived by merging microfluidics with transformation polaritonics. Our findings provide a new pathway for polariton manipulation, bearing great potential in polaritonic metasurfaces[41] and on-chip polaritonic elements and circuits[42]. The rich synergies between polaritonics and microfluidics could render our platform broadly applicable in molecular sensing, chemical reaction monitoring, and biological transportation.

**Acknowledgement**

The project was sponsored by the National Natural Science Foundation of China (Grant Nos. 62305288, 61975176), the Key Research and Development Program of the Ministry of Science and Technology (Grant Nos. 2022YFA1404704, 2022YFA1405200, and 2022YFA1404902), the Key Research and Development Program of Zhejiang Province (Grant No. 2022C01036), and the Fundamental Research Funds for the Central Universities. A Project Supported by Scientific Research Fund of Zhejiang Provincial Education Department (Grant No. Y202353629).

**Conflict of interests**

The authors declare no conflict interests.

**Author Contributions**

H.C. and Y.W. convinced the idea. Z.C. and Y.W. performed theoretical analysis and numerical simulations. All the authors have analyzed and discussed the results. Z.C. and Y.W. co-wrote the manuscript with inputs from all the other authors. H.C. and Y.W. supervised the project.

**References**

1. Basov, D. N.; Fogler, M. M.; Garcia de Abajo, F. J. Polaritons in van der Waals materials. *Science* **2016,** 354, (6309), aaag1992.
2. Low, T.; Chaves, A.; Caldwell, J. D.; Kumar, A.; Fang, N. X.; Avouris, P.; Heinz, T. F.; Guinea, F.; Martin-Moreno, L.; Koppens, F. Polaritons in layered two-dimensional materials. *Nat. Mater.* **2017,** 16, (2), 182-194.
3. Zhang, Q.; Hu, G.; Ma, W.; Li, P.; Krasnok, A.; Hillenbrand, R.; Alù, A.; Qiu, C.-W. Interface nano-optics with van der Waals polaritons. *Nature* **2021,** 597, (7875), 187-195.

4. Wu, Y.; Duan, J.; Ma, W.; Ou, Q.; Li, P.; Alonso-González, P.; Caldwell, J. D.; Bao, Q. Manipulating polaritons at the extreme scale in van der Waals materials. *Nat. Rev. Phys.* **2022,** 4, (9), 578-594.

5. Duan, J.; Chen, R.; Li, J.; Jin, K.; Sun, Z.; Chen, J. Launching phonon polaritons by natural boron nitride wrinkles with modifiable dispersion by dielectric environments. *Adv. Mater.* **2017,** 29, (38), 1702494.

6. Fali, A.; White, S. T.; Folland, T. G.; He, M.; Aghamiri, N. A.; Liu, S.; Edgar, J. H.; Caldwell, J. D.; Haglund, R. F.; Abate, Y. Refractive index-based control of hyperbolic phonon-polariton propagation. *Nano Lett.* **2019,** 19, (11), 7725-7734.

7. Dai, S.; Quan, J.; Hu, G.; Qiu, C. W.; Tao, T. H.; Li, X.; Alu, A. Hyperbolic phonon polaritons in suspended hexagonal boron nitride. *Nano Lett.* **2019,** 19, (2), 1009-1014.

8. Hu, H.; Yu, R.; Teng, H.; Hu, D.; Chen, N.; Qu, Y.; Yang, X.; Chen, X.; McLeod, A. S.; Alonso-Gonzalez, P.; Guo, X.; Li, C.; Yao, Z.; Li, Z.; Chen, J.; Sun, Z.; Liu, M.; Garcia de Abajo, F. J.; Dai, Q. Active control of micrometer plasmon propagation in suspended graphene. *Nat. Commun.* **2022,** 13, (1), 1465.

9. Zentgraf, T.; Liu, Y.; Mikkelsen, M. H.; Valentine, J.; Zhang, X. Plasmonic Luneburg and Eaton lenses. *Nat. Nanotechnol.* **2011,** 6, (3), 151-155.

10. Duan, J.; Álvarez-Pérez, G.; Tresguerres-Mata, A. I. F.; Taboada-Gutiérrez, J.; Voronin, K. V.; Bylinkin, A.; Chang, B.; Xiao, S.; Liu, S.; Edgar, J. H.; Martin, J. I.; Volkov, V. S.; Hillenbrand, R.; Martín-Sánchez, J.; Nikitin, A. Y.; Alonso-González, P. Planar refraction and lensing of highly confined polaritons in anisotropic media. *Nat. Commun.* **2021,** 12, (1), 4325.

11. Deng, F.; Guo, Z.; Ren, M.; Su, X.; Dong, L.; Liu, Y.; Shi, Y.; Chen, H. Bessel beam generated by the zero-index metalens. *Prog. Electromagn. Res.* **2022,** 174, 89-106.

12. Folland, T. G.; Fali, A.; White, S. T.; Matson, J. R.; Liu, S.; Aghamiri, N. A.; Edgar, J. H.; Haglund, R. F., Jr.; Abate, Y.; Caldwell, J. D. Reconfigurable infrared hyperbolic metasurfaces using phase change materials. *Nat. Commun.* **2018,** 9, (1), 4371.

13. Chaudhary, K.; Tamagnone, M.; Yin, X.; Spagele, C. M.; Oscurato, S. L.; Li, J.; Persch, C.; Li, R.; Rubin, N. A.; Jauregui, L. A.; Watanabe, K.; Taniguchi, T.; Kim, P.; Wuttig, M.; Edgar, J. H.; Ambrosio, A.; Capasso, F. Polariton nanophotonics using phase-change materials. *Nat. Commun.* **2019,** 10, (1), 4487.

14. Aghamiri, N. A.; Hu, G.; Fali, A.; Zhang, Z.; Li, J.; Balendhran, S.; Walia, S.; Sriram, S.; Edgar, J. H.; Ramanathan, S.; Alu, A.; Abate, Y. Reconfigurable hyperbolic polaritonics with correlated oxide metasurfaces. *Nat. Commun.* **2022,** 13, (1), 4511.

15. Haynes, W. M., *Handbook of Chemistry and Physics*. CRC Press: 2016.

16. Wang, H.; Janzen, E.; Wang, L.; Edgar, J. H.; Xu, X. G. Probing mid-infrared phonon polaritons in the aqueous phase. *Nano Lett.* **2020,** 20, (5), 3986-3991.

17. Virmani, D.; Bylinkin, A.; Dolado, I.; Janzen, E.; Edgar, J. H.; Hillenbrand, R. Amplitude- and phase-resolved infrared nanoimaging and nanospectroscopy of polaritons in a liquid environment. *Nano Lett.* **2021,** 21, (3), 1360-1367.

18. O'Callahan, B. T.; Park, K. D.; Novikova, I. V.; Jian, T.; Chen, C. L.; Muller, E. A.; El-Khoury, P. Z.; Raschke, M. B.; Lea, A. S. In liquid infrared scattering scanning near-field optical microscopy for chemical and biological nanoimaging. *Nano Lett.* **2020,** 20, (6), 4497-4504.

19. Zhao, C.; Liu, Y.; Zhao, Y.; Fang, N.; Huang, T. J. A reconfigurable plasmofluidic lens. *Nat. Commun.* **2013,** 4, 2305.

20. Psaltis, D.; Quake, S. R.; Yang, C. Developing optofluidic technology through the fusion of microfluidics and optics. *Nature* **2006,** 442, (7101), 381-386.
21. Monat, C.; Domachuk, P.; Eggleton, B. J. Integrated optofluidics: A new river of light. *Nat. Photonics* **2007,** 1, (2), 106-114.
22. Juan, M. L.; Righini, M.; Quidant, R. Plasmon nano-optical tweezers. *Nat. Photonics* **2011,** 5, (6), 349-356.
23. Yang, Y.; Liu, A. Q.; Chin, L. K.; Zhang, X. M.; Tsai, D. P.; Lin, C. L.; Lu, C.; Wang, G. P.; Zheludev, N. I. Optofluidic waveguide as a transformation optics device for lightwave bending and manipulation. *Nat. Commun.* **2012,** 3, 651.
24. Yan, R.; Yang, Y.; Tu, X.; Huang, T.; Liu, Y.; Song, C. Optofluidic light routing via analytically configuring streamlines of microflow. *Microfluidics and Nanofluidics* **2019,** 23, (8), 101.
25. Liu, H. L.; Zuo, Y. F.; Zhu, X. Q.; Yang, Y. Optofluidic gradient refractive index resonators using liquid diffusion for tunable unidirectional emission. *Lab Chip* **2020,** 20, (15), 2656-2662.
26. Li, Q.; van de Groep, J.; White, A. K.; Song, J. H.; Longwell, S. A.; Fordyce, P. M.; Quake, S. R.; Kik, P. G.; Brongersma, M. L. Metasurface optofluidics for dynamic control of light fields. *Nat. Nanotechnol.* **2022,** 17, (10), 1097-1103.
27. Morgado, T. A.; Silveirinha, M. G. Negative landau damping in bilayer graphene. *Phys. Rev. Lett.* **2017,** 119, (13), 133901.
28. Morgado, T. A.; Silveirinha, M. G. Drift-induced unidirectional graphene plasmons. *ACS Photonics* **2018,** 5, (11), 4253-4258.
29. Correas-Serrano, D.; Gomez-Diaz, J. S. Nonreciprocal and collimated surface plasmons in drift-biased graphene metasurfaces. *Phys. Rev. B* **2019,** 100, (8), 081410.
30. Dong, Y.; Xiong, L.; Phinney, I. Y.; Sun, Z.; Jing, R.; McLeod, A. S.; Zhang, S.; Liu, S.; Ruta, F. L.; Gao, H.; Dong, Z.; Pan, R.; Edgar, J. H.; Jarillo-Herrero, P.; Levitov, L. S.; Millis, A. J.; Fogler, M. M.; Bandurin, D. A.; Basov, D. N. Fizeau drag in graphene plasmonics. *Nature* **2021,** 594, (7864), 513-516.
31. Zhao, W.; Zhao, S.; Li, H.; Wang, S.; Wang, S.; Utama, M. I. B.; Kahn, S.; Jiang, Y.; Xiao, X.; Yoo, S.; Watanabe, K.; Taniguchi, T.; Zettl, A.; Wang, F. Efficient Fizeau drag from Dirac electrons in monolayer graphene. *Nature* **2021,** 594, (7864), 517-521.
32. Khorasaninejad, M.; Chen, W. T.; Oh, J.; Capasso, F. Super-dispersive off-axis meta-lenses for compact high resolution spectroscopy. *Nano Lett.* **2016,** 16, (6), 3732-3737.
33. Chen, Q.; Jian, A.; Li, Z.; Zhang, X. Optofluidic tunable lenses using laser-induced thermal gradient. *Lab Chip* **2016,** 16, (1), 104-111.
34. Zentgraf, T.; Valentine, J.; Tapia, N.; Li, J.; Zhang, X. An optical "Janus" device for integrated photonics. *Adv. Mater.* **2010,** 22, (23), 2561-2564.
35. Huidobro, P. A.; Nesterov, M. L.; Martin-Moreno, L.; Garcia-Vidal, F. J. Transformation optics for plasmonics. *Nano Lett.* **2010,** 10, (6), 1985-1990.
36. Liu, Y.; Zentgraf, T.; Bartal, G.; Zhang, X. Transformational plasmon optics. *Nano Lett.* **2010,** 10, (6), 1991-1997.
37. Kadic, M.; Guenneau, S.; Enoch, S.; Huidobro, P. A.; Martín-Moreno, L.; García-Vidal, F. J.; Renger, J.; Quidant, R. Transformation plasmonics. *Nanophotonics* **2012,** 1, (1), 51-64.
38. Song, L.; Shen, L.; Wang, H. Squeezing of hyperbolic polaritonic rays in cylindrical lamellar structures. *Prog. Electromagn. Res.* **2022,** 174, 23-32.
39. Mei, Z. L.; Cui, T. J. Arbitrary bending of electromagnetic waves using isotropic materials.

*Journal of Appl. Phys.* **2009,** 105, (10), 104913.

40. Rahm, M.; Roberts, D. A.; Pendry, J. B.; Smith, D. R. Transformation-optical design of adaptive beam bends and beam expanders. *Opt. Express* **2008,** 16, (15), 11555-11567.

41. Ding, F. A review of multifunctional optical gap-surface plasmon metasurfaces. *Prog. Electromagn. Res.* **2022,** 174, 55-73.

42. Yao, D. Y.; He, P. H.; Zhang, H. C.; Zhu, J. W.; Hu, M.; Cui, T. J. Miniaturized photonic and microwave integrated circuits based on surface plasmon polaritons. *Prog. Electromagn. Res.* **2022,** 175, 105-125.

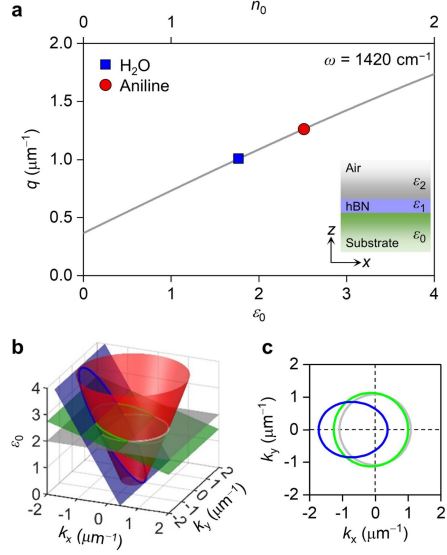

Figure 1. Breaking polariton propagation symmetry by dielectric environment gradients. (a) Dispersion relation of PhPs in hBN as a function of the permittivity of the substrate ($\varepsilon_0$) at the frequency ($\omega$) of 1420 cm$^{-1}$. The superstrate is air ($\varepsilon_2 = 1$) and the thickness of hBN was set as 100 nm. Inset is the illustration of the three-layer structure. (b) Corresponding isofrequency surface (red surface) of PhPs in hBN with varied $\varepsilon_0$. Gray surface is the slice at $\varepsilon_0 = 2$, whereas blue and green surfaces correspond to $\partial \varepsilon_0/\partial k_x = -1.86$ and $-0.33$, respectively. (c) Projections of intersections between slices and the isofrequency surface in b.

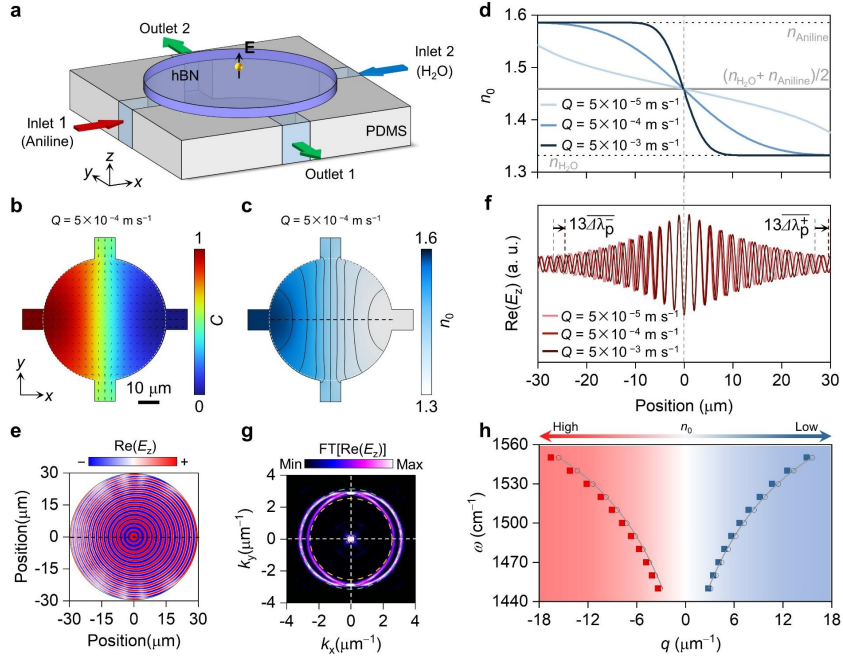

Figure 2. Dragging polaritons with microfluidics. (a) Illustration of the microfluidic structures for polariton drag. (b) Normalized concentration distribution over the microfluidic chamber. Black arrows indicate the flow direction. White dashed circle marks the chamber. (c) Corresponding refractive index distribution. Solid curves mark refractive index contours. (d) Line scan profiles along the black dashed line in c under different flow rates. (e) Simulated electric field distribution under the flow rate of $5\times10^{-4}$ m s$^{-1}$ at 1450 cm$^{-1}$. (f) Corresponding line scan profiles across the chamber center as indicated by the black dashed lines in e. Gray curve represents the line scan signals simulated using the uniform aniline and water mixture (gray line in d). (g) Fourier transform of the electric field in e. Outer dashed circle is the IFC with $n_0 = 1.586$, while the inner one is the IFC with $n_0 = 1.332$. (h) Dispersion relation of polaritons with (squares) and without (circles) refractive index gradients. In simulations the flow rate was set as $5\times10^{-3}$ m s$^{-1}$. Gray curves are dispersion relations calculated using the uniform aniline and water mixture.

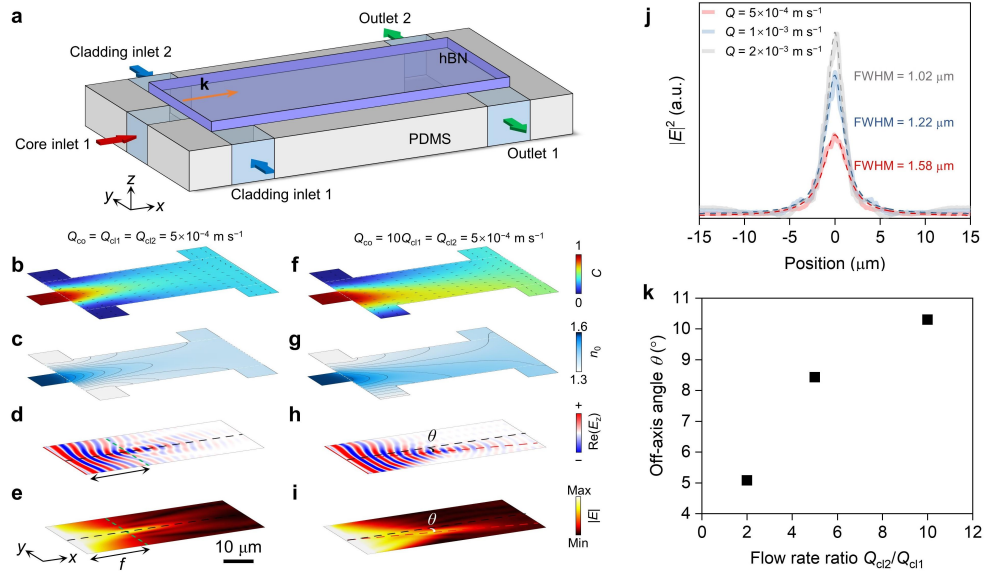

Figure 3. Planar lens and off-axis lens for polariton subwavelength focusing. (a) Illustration of the microfluidic structures for polariton focusing. Aniline and water are respectively core and cladding fluids. (b) Normalized concentration distribution. (c) Corresponding refractive index distribution. (d,e) Simulated electric field distribution over the rectangular chamber at 1420 cm$^{-1}$. Green dashed lines indicate the focal position. (f–i) Same to b–e but for asymmetric flows. Red dashed lines represent the focal lines with an off-axis angle of $\theta$. (j) Extracted line scan profiles along the green dash lines in d and e. Dashed curves are Lorentzian fittings. (k) Off-axis angles as a function of the flow rate ratio.

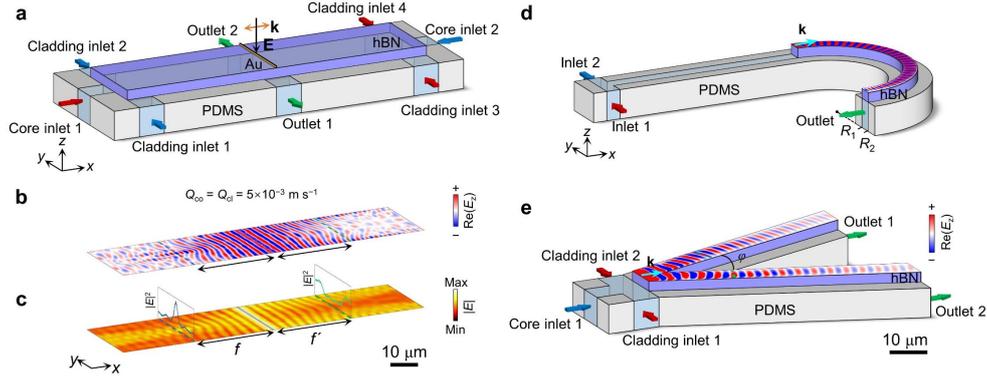

Figure 4. Polaritonic Janus lens, bend and splitter based on microfluidics. (a) Illustration of the Janus lens. Red and blue arrows represent the aniline and $H_2O$ flows. (b) Simulated field distribution at 1420 cm$^{-1}$, indicating converged and diverged wavefronts in the −x and +x directions. (c) Corresponding field distribution. Insets are the line scan profiles along the blue and green dashed lines. (d) Illustration of the bend for polariton rotation. Red and blue arrows represent the aniline and $H_2O$ flows with the flow rates of 0.131 and 0.111 m s$^{-1}$, respectively. The inner ($R_1$) and outer ($R_2$) radii of the semicircular channel are 20 and 23.81 μm. Simulated field distribution is overlaid on the top surface of hBN. (e) Illustration of the Y-shaped polariton splitter with a split angle ($\varphi$) of 37.5°. Aniline and $H_2O$ serve as core and cladding flows with the flow rates of $2\times10^{-3}$ and $1\times10^{-2}$ m s$^{-1}$, respectively.